# Electrochemical characterization of a Fe-based shape memory alloy in an alkaline medium and the behaviour in aggressive conditions

A. Collazo, R. Figueroa, C. Mariño-Martínez, X.R. Nóvoa, C. Pérez[*]

*ENCOMAT Group, CINTECX, Universidade de Vigo, Vigo 36310, Spain*

ABSTRACT

A Fe-17Mn-6Si-19Cr-4Ni-1(V,C) shape memory steel (SMS) was characterised electrochemically in its unstrained and pre-strained conditions. The work focused on analysing the passive films generated in alkaline conditions, and on the behaviour of those passive samples in various $Cl^-/OH^-$ ratios. The passive films were developed by cyclic voltammetry and their characterisation was performed by electrochemical impedance spectroscopy (EIS) and X-ray photoelectron spectroscopy (XPS); these tests were also carried out in 304 L stainless steel specimens for comparison purposes. The results indicated that the film characteristics (thickness and composition) were similar in all the samples, although higher corrosion resistance was observed in the 304 L due to the higher Cr content. In addition, the behaviour of the passive samples in aggressive conditions was assessed by potentiodynamic measurements. The results stated that the shape memory steels were more sensitive than the stainless steel to the aggressive conditions, especially when the alloy was pre-strained. It was also observed that the SMS exhibited a characteristic corrosion morphology localized at the grain boundaries.

## 1. Introduction

Shape memory alloys (SMAs) are a variety of smart materials characterized by having Shape Memory Effect (SME), which consists of the capability to recover its original shape through temperature variation once the material has been deformed beyond its elastic limit [1]. Since the discovery of the SME in the Fe-Mn-Si shape memory alloys by Sato et al. in 1982 [2], they have been widely studied as an alternative to the more traditional NiTi alloys due to their low cost, uncomplicated manufacturing process and better mechanical properties (higher stiffness and strength). The SME in the Shape Memory Steels (SMS) is based on the stress-induced transformation from the parent γ-austenite phase (fcc, face-centred cubic) to the ε-martensite phase (hcp, hexagonal close-packed) and the reverse transformation (ε→γ) by heating. These structural modifications are associated with the corresponding dimensional changes [3,4].

In recent years, the application of SMS in civil engineering has gained attraction, mainly for constrained recovery applications such as pipe joints, rail couplings or pre-stressing reinforcement elements; this latter represents a most promising field of application. In this, the SMS component is pre-strained before being embedded in concrete. Once the concrete has been hardened, the SMS is heated to activate the reverse transformation; however, its deformation is constrained in the concrete, and tensile stress is generated in the SMS (*recovery stress*). Consequently, compressive stress is developed in the concrete. The advantage of this procedure is that no pre-stressing tools, such as mechanical jacks and anchor heads, are needed since the pre-stressing is achieved by the SME of the alloy [5,6]. However, these Fe-Mn-Si-based SMAs exhibit poor corrosion resistance and limited SME, restricting their use in engineering applications. In the last decades, new formulations with additional allowing elements were developed to overcome these drawbacks. In this context, adding Cr and/or Ni improves the corrosion resistance of these alloys maintaining good SME [7,8].

Moreover, it was found that the precipitation of fine particles of NbC [9], VN [10] or VC [11] in Fe-Mn-Si-based alloys remarkably improved their SME. However, these studies focused on the structural and mechanical characterization of the new alloys, and only a few addressed the corrosion behaviour [8,12–14]. As afore mentioned, this aspect is essential in applications based on pre-stressing reinforcement elements. Lee et al. [14] studied the corrosion behaviour of a shape memory steel in concrete, simulating pore solutions, using potentiodynamic techniques. They corroborate its good corrosion resistance in alkaline conditions, although it was susceptible to pitting corrosion in the presence of chloride ions. Nevertheless, more detailed studies are needed to understand the behaviour of these alloys embedded in concrete.

The present work deals with the electrochemical survey of a Fe-17Mn-6Si-19Cr-4Ni-1(V,C) shape memory alloy in alkaline conditions, close to those inside the concrete pores, considering not only unstrained but also pre-stained states since this is the condition in which the SMS will be embedded in the concrete. The characterization of the passive film generated in alkaline conditions was performed by electrochemical impedance spectroscopy (EIS) and X-ray photoelectron spectroscopy (XPS). In addition, the response of the previously passivated samples in aggressive conditions, such as the presence of chlorides or the alkalinity decrease due to the carbonation process was also studied.

## 2. Experimental

### 2.1. Materials

The SMS specimens were rebars with a diameter of 11 mm, provided by the "RE-FER" company (https://www.re-fer.eu/) in two states: unstrained and 5% pre-strained. For comparative purposes, the electrochemical experiments were also performed in AISI 304 L stainless steel rebars with a diameter of 14 mm. The chemical composition of both types of samples is given in Table 1.

The specimens were mechanically grounded with successive grades of SiC papers up to 1200 grit. They were then polished with diamond paste of 1 μm grain size to obtain mirror-like appearance. Finally, samples were ultrasonically cleaned before immersion in the different solutions test.

### 2.2. Morphological characterization

The surface morphology of the SMS samples was studied by scanning electron microscopy (SEM). The equipment employed was an Electroscan JSM-54® model from OXFORD® instruments using an acceleration voltage of 20 kV. The X-ray diffraction (XRD) technique was used to identify the phases present in both states. The equipment employed was a Siemens® D5000 powder diffractometer with a monochromatic CuK$_\alpha$ radiation, $\lambda_\alpha$ = 1.5 Å; the step size and time were 0.02° and 4 s, respectively.

The analytical characterisation of the passive films generated in the alkaline medium was assessed by X-ray photoelectron spectroscopy (XPS) using a Thermo Scientific K-Alpha ESCA® instrument equipped with Al K$_\alpha$ monochromatic radiation (1486.6 eV). The measurement was done in a Constant Analyser Energy mode (CAE) with a 100 eV pass energy for survey spectra and 20 eV pass energy for high-resolution spectra.

### 2.3. Electrochemical measurements

The passive films were generated by cyclic voltammetry (CV), submitting the samples to six cycles to obtain a film thick enough to be analysed. The electrolyte was a NaOH 0.1M + KOH 0.1 M solution (pH=13) close to the concrete pores' solution. The potential was scanned from hydrogen to oxygen evolution reactions at dE/dt = 5 mV·s$^{-1}$.

The characterisation of the semiconducting properties of the films generated was performed from the Mott-Schottky approach. The relation between the space charge capacitance ($C_{SC}$) and the potential is given by [15,16]:

$$\frac{1}{C_{SC}^2} = \frac{\pm 2}{N e \varepsilon_r \varepsilon_o}\left(E - E_{fb} - \frac{kT}{e}\right) \quad (1)$$

where the positive sign is for n-type semiconductors and the negative for p-type semiconductors. N is the dopant concentration (cm$^{-3}$), $e$ is the electron charge (1.6 × 10$^{-19}$C), $\varepsilon_o$ is the vacuum permittivity (8.85×10$^{-14}$ F·cm$^{-1}$), $\varepsilon_r$ is the relative dielectric constant of the passive film (taken as 15.6 in the present study [15]), E$_{fb}$ is the flat band potential, k is the Boltzmann constant (1.38×10$^{-23}$ J·k$^{-1}$) and T the absolute temperature. N can be determined from the magnitude of the slope of the experimental $C_{SC}^{-2}$ vs E plots, while the slope sign identifies the type of semiconductor film. The thickness of the space-charge layer, L$_{SC}$, can be calculated from Eq. (2) [17]:

$$L_{SC} = \left[\frac{\pm 2\varepsilon_r\varepsilon_o}{Ne}\left(E - E_{fb} - \frac{kT}{e}\right)\right]^{0.5} \quad (2)$$

Where the symbols have the same meaning as in Eq. (1).

The measured capacitance, C, was obtained from Eq. (3):

$$C = \frac{1}{\omega Z''} \quad (3)$$

where Z'' is the imaginary component of the impedance and $\omega = 2\pi f$ is the angular frequency. The M-S approach assumes that the space charge capacitance is much smaller than the Helmholtz double-layer capacitance. Since both are in a series arrangement, the measured capacitance will equal $C_{SC}$.

The impedance measurements were conducted after 24 h of the passive layer generation by sweeping the potential anodically in the passivity region, from −0.4 to 0.2 V$_{vsHg/HgO}$ with a step of 25 mV, at 1 kHz frequency and using an AC signal with 10 mV$_{rms}$ amplitude.

After the Mott-Schottky tests, impedance measurements in a broader frequency range were carried out to investigate the processes at the metal/passive film/solution interfaces. The measurements were also done in the passivity region, every 50 mV in the anodic direction, the frequency range was from 10 kHz to 1 mHz using 7 points per decade and an AC signal with 10 mV$_{rms}$ amplitude.

The corrosion behaviour of the specimens in aggressive environments was assessed using the CV technique. The previous step was the passive film generation employing the procedure mentioned above. Once the film had been generated, the samples remained in the alkaline solution for 24 h, after which this electrolyte was removed, and the cell was filled with the new one. Different solutions were tested, in which a 0.1 M NaCl concentration was kept constant, and the pH was lowered at 0.5 pH unit steps by diluting the original NaOH 0.1M + KOH 0.1 M solution. This experimental procedure aimed to simulate the loss of alkalinity that occurs in the reinforced concrete due to the carbonation process. The voltammetry was performed after one hour of stabilization time. The potential was scanned from −50 mV cathodic respect to the OCP up to oxygen evolution reaction or when high current densities were recorded (in the order of mA·cm$^{-2}$) at 1 mV·s$^{-1}$ scan rate.

A conventional three-electrode arrangement was used to perform all the electrochemical experiments. The metallic specimens were the working electrode defining an area of 0.33 cm$^2$, and a large platinum mesh was used as the counter electrode. Two reference electrodes were employed depending on the electrolyte. A Hg/HgO 0.1 M KOH electrode was used with NaOH 0.1M + KOH 0.1 M solution during the generation and characterization of the passive film to avoid contamination with chloride ions. The saturated calomel electrode (SCE) was employed in chloride-containing electrolytes experiments.

An Autolab 30 Potentiostat with an FRA module, from EcoChemie®, was employed to perform all the electrochemical measurements.

**Table 1**
Chemical composition (wt%) of the SMS and the AISI 304 L SS.

| Alloy | Mn | Si | Cr | Ni | Mo | V | C | Fe |
|---|---|---|---|---|---|---|---|---|
| SMS | 17.17 | 5.06 | 9.99 | 4.21 | — | 0.74 | 0.18 | Bal. |
| AISI 304L | 1.79 | 0.34 | 18.22 | 8.58 | 0.43 | — | 0.023 | Bal. |

## 3. Results and discussion

### 3.1. Morphological characterization

Fig. 1 displays the microstructure observed for unstrained and pre-strained SMS samples. The γ-austenite phase is the majority; regardless of the SMS state, the average grain size can be estimated at around 17 μm. Also, fine ε-martensite plates can be appreciated, more evident in the pre-strained specimens. The presence of this phase in the unstrained samples is unexpected; however, it was also observed by other researchers, and it could be due to the internal stress around the VC nanoparticles present in this type of SMS [3]. The pre-strained samples also exhibit twins generated during the mechanical stressing. In addition, it highlights the presence of an additional phase located at grain boundaries, which may be ascribed to some residual high-temperature ferrite (δ-ferrite). It is well-documented that austenitic stainless steels, depending on the balance of ferrite-stabilizers (Cr, Si or V) to austenite-stabilizers (Ni, C or Mn) elements, can primarily solidify as δ-ferrite due to the segregation of these ferrite-promoting elements. Usually, it is present in low volume fraction (less than 2%) [18,19]. Several researchers [20,21] corroborated the presence of this phase in the shape memory alloys with chemical compositions close to that of the SMS studied in the present work.

The XRD technique was employed for assessing the identification of the phases present. Fig. 2 displays a detail of the obtained diffractograms. As expected, the diffraction peaks correspond to γ-austenite and ε-martensite phases, no additional peaks that could be ascribed to δ-ferrite are observed, probably due to its small volume fraction. On the other hand, the γ/ε ratio decreases in the pre-strained SMS samples, which corroborates the stress-induced martensitic transformation (γ → ε) [22]. The pre-strained SMS was thermally activated by heating at a temperature of 160 °C for 15 min [14]; the recorded XRD pattern is also depicted in Fig. 2, it can be observed an increase in the γ/ε ratio that confirms the reverse transformation (ε → γ).

### 3.2. Characterization of the passive film generated in the alkaline medium

Cyclic voltammetry (CV) is a common technique for characterizing the passive layers generated on metallic substrates; it provides a good understanding of the oxidation and reduction processes occurring on passive metals. The CVs performed in the SMS samples (both states) immersed in 0.1 M NaOH+ 0.1 M KOH solution are shown in Fig. 3. For comparative purposes, it also includes the cyclic voltammogram recorded for AISI 304 L. This material has been studied extensively in alkaline conditions [23,24]. Some characteristic peaks are observed in both types of samples, although significant differences are evident, which should be related to the presence of manganese. The more significant peaks were interpreted based on the literature [23–27]. It may be summarised in the following main features: for AISI 304 L, the most cathodic region, from hydrogen evolution up to about $-0.3$ $V_{vsHg/HgO}$, is related to the Fe redox processes, $Fe/Fe^{2+}$, $Fe^{2+}/Fe_3O_4$ and $Fe_3O_4/Fe^{3+}$, corresponding the peak centred at $-0.7$ $V_{vsHg/HgO}$ in the forward scan to the magnetite formation. In the same region, for the SMS, some Mn redox processes, specifically $Mn/Mn^{2+}$ and $Mn^{2+}/Mn_3O_4$, must be considered [27], in addition to the Fe activity. Highlights the lower current density recorded at the magnetite formation peak. Although the reason is unclear, it could be related to the surface coverage with manganese oxides that partially block the iron activity. At more anodic potentials, between $-0.4$ $V_{vsHg/HgO}$ to about $+0.2$ $V_{vsHg/HgO}$, a passivity region is observed in all specimens, characterized by a plateau zone with current densities in the order of 10–15 μA·cm$^2$. At higher potentials, additional redox activity is noticed; in the case of AISI 304 L, a broad peak centred about $+0.4$ $V_{vsHg/HgO}$ accounts for the $Cr^{3+}/Cr^{6+}$ and $Ni^{2+}/Ni^{3+}$ redox processes. Again, a distinct behaviour is observed in the SMS, the current densities are higher, and two peaks are distinguished. The first peak can be interpreted considering not only the Cr and Ni activity but also the $Mn_3O_4/Mn^{3+}$ transformation, and the peak recorded above 0.5 $V_{vsHg/HgO}$ (close to the $O_2$ evolution) is ascribed to the $Mn^{3+}/Mn^{4+}$ process with the formation of $MnO_2$ [27].

Analytical information on the generated passive films has been extracted from the XPS measurements. The acquired survey spectra are depicted in Fig. 4(a). The peaks corresponding to the main alloying elements, Cr, Ni, Si and Mn, are distinguished, together with the Fe, O and C signals. The high-resolution spectra recorded for Fe 2p, Mn 2p and Cr 2p are shown in Fig. 4(b), (c) and (d). The unambiguous assignation of the oxidation states of Fe and Mn is not easy, owing to the superposition of the binding energy for the $Fe^{3+}$ (710.3–711.5 eV) and $Fe^{2+}$ (709.3–710.9 eV) oxides and the $Mn^{4+}$ (641.6–642.8 eV), $Mn^{3+}$ (641.2–641.7 eV) and $Mn^{2+}$ (640.4–641.5 eV) oxides [28]. Despite that and based on previous research [29,30], an adscription to the observed peaks is proposed. The Fe $2p_{3/2}$ signal presents a single peak at 710.7 eV that can assign to $Fe^{3+}$ species (oxides/hydroxides); this adscription is corroborated by the presence of the satellite signal at 718.8 eV. The Mn $2p_{3/2}$ signal exhibits a maximum at 641.57 eV that may correspond to $Mn^{3+}$. A shoulder at higher energy, 642.5 eV, is ascribed to $Mn^{4+}$; at lower energies (640.0 eV), a small shoulder seems to appear corresponding to the $Mn^{2+}$ oxidation state; this last adscription is supported by the presence of a satellite peak at 645 eV due to MnO. As expected, the Mn peak's intensity is markedly higher for the SMS owing to the higher Mn content. The Cr $2p_{3/2}$ signal is characterized by an intense peak at 576.8 eV ascribed to $Cr^{3+}$ and maybe a small shoulder at 574.5 eV that should correspond to $Cr^0$.

Fig. 4(e) shows the obtained depth profiles, following the Fe 2p and O 2p signals. The film thickness is about 20 nm for all samples, ranging between 19 nm for AISI 304 L and 23 nm for s-SMS. These values agree with those obtained in previous studies for films generated

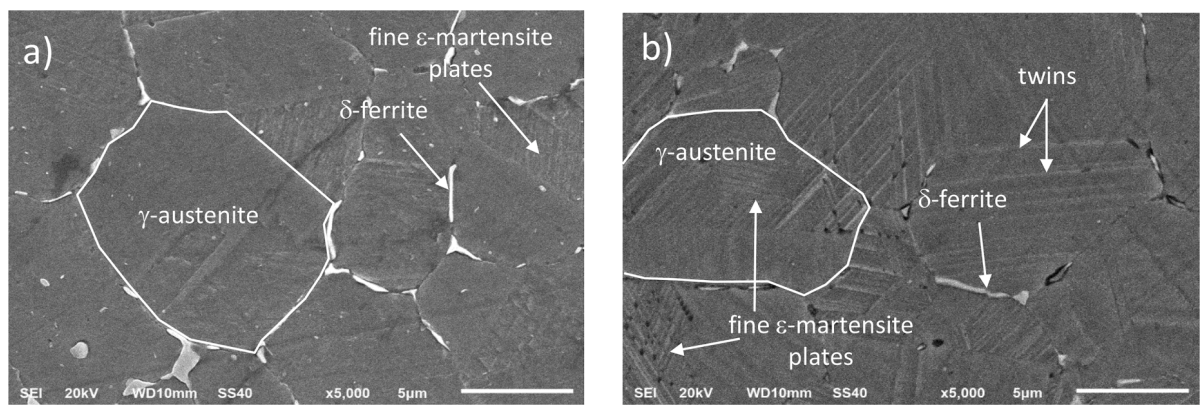

**Fig. 1.** SEM micrographs corresponding to the unstrained (a) and pre-strained (b) shape memory specimens in the as-received states.

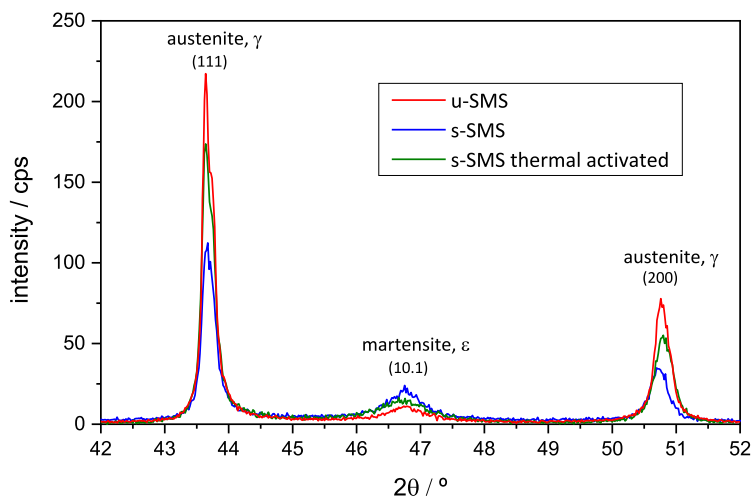

**Fig. 2.** A detail of the XRD patterns of the unstrained, pre-strained and thermal-activated SMSs samples.

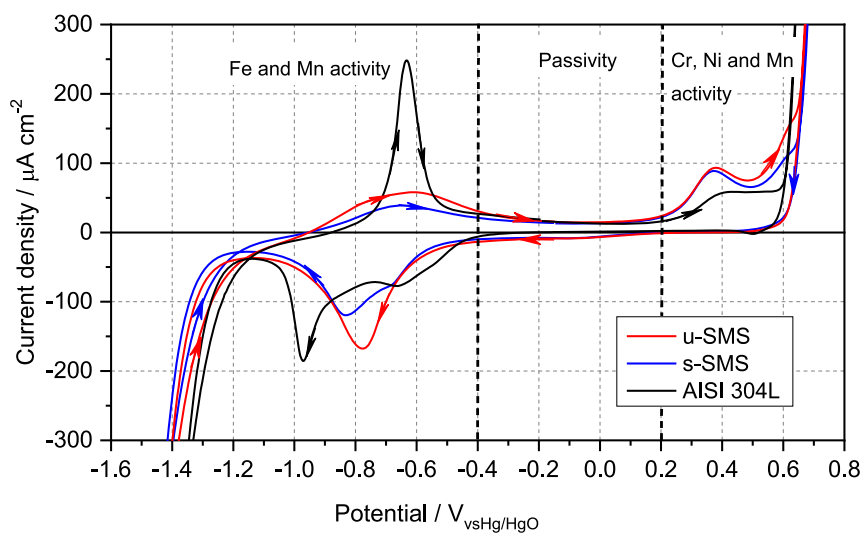

**Fig. 3.** Cyclic voltammograms (6th cycle) for unstrained (u-SMS) and pre-strained (s-SMS) shape memory steel and AISI 304 L performed at 5mV·s$^{-1}$ scan rate in 0.1 M NaOH + 0.1 M KOH solution.

potentiodynamically [23].

The semiconductive nature of the generated passive films was characterised based on the Mott-Schottky (M-S) analysis. The $C_{SC}^{-2}$ vs applied potential profiles (M-S plots) are shown in Fig. 5(a). The applied potential window was confined to the passivity region, from −0.4 to +0.2 V$_{vsHg/HgO}$, where the film is stable, and the dopant concentration could be considered independent of the voltage [16].

Three different regions are distinguished. At low potentials, the oxide films behave as an n-type semiconductor due to oxygen vacancies and/or Fe or Mn cation interstitials., A plateau region is observed at higher potentials, which may be interpreted as dielectric behaviour [31]. Finally, at more anodic potentials, the M-S behaviour can be associated with a p-type semiconduction, probably owing to cations vacancies from the chromium oxide. Other researchers have also reported this duplex semiconductive character depending on the applied potential for stainless steels [32–35] and some shape memory steels [36, 37]. However, in the studied SMS, the p-type response is not as evident as in the AISI 304 L; therefore, the extracted values should be considered qualitative or semiquantitative.

The donor ($N_D$) and acceptor ($N_A$) dopant concentrations can be assessed from the slope of the M-S plots. Table 2 shows the obtained results, which are around 10$^{20}$–10$^{21}$ cm$^{-3}$, regardless of the material.

These values agree with those reported in the literature [34,36].The higher $N_D$ values for the AISI 304 L suggest a highly disordered amorphous layer. Di Paola [31] and Azumi et al. [17] pointed out that thin films contain high donor defect concentration, and their density decreases with the increase of the film thickness because the layers tend to adopt a more stable structure as the thickness increases. Based on that, the outer layer (Fe and Mn oxides) seems thinner in the 304 L than in the SMS. On the other hand, the $N_A$ densities are similar in all the samples, even though the AISI 304 L seems to generate a less defective Cr-oxide layer.

The evolution of the space charge thickness ($L_{SC}$) with the applied potential is illustrated in Fig. 5(b). It may be understood based on the model proposed by Hakiki et al. for the passive films formed on Fe-Cr alloys [38], which consists of the coexistence of an inner region, in contact to the metal, essentially made by chromium oxide and with p-type behaviour, and an outer region mainly composed by iron oxides with n-type behaviour. Thus, the presence of two space charge regions developed in the passive films [35] seems reasonable. In the more cathodic region, the $L_{SC}$ increases with the applied potential suggesting the outer layer thickening owing to the Fe and Mn oxidation. In addition, the $L_{SC}$ values are slightly higher for s-SMS, in good agreement with the oxide film thicknesses measured by XPS. However, at the more anodic potentials, the opposite evolution is observed; the $L_{SC}$ decreases as the applied potential increases. Although the reason is unclear, it may be related to the $Cr^{3+}/Cr^{6+}$ oxidation process, which shows the peak activity at 0.4 $V_{vsHg/HgO}$ (see Fig. 3). The $L_{SC}$ decrease is less pronounced for SMS maybe owing to the $Mn_3O_4/Mn^{3+}$ transformation that behaves like as Fe species (n-type behaviour) and tends to a film thickening.

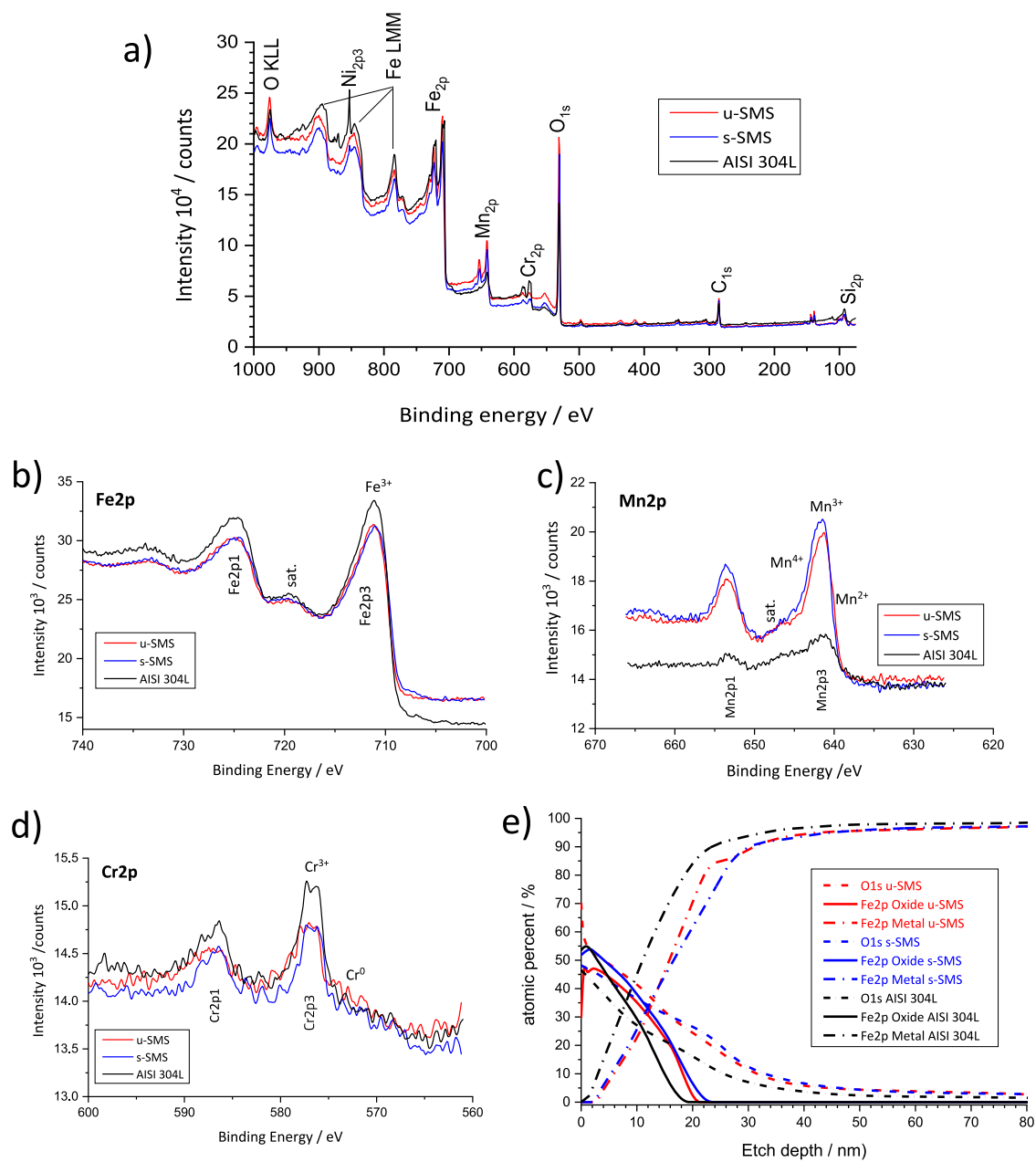

**Fig. 4.** XPS survey spectra (a), high-resolution spectra of Fe 2p (b), Mn 2p (c) Cr 2p (d), and XPS depth profiles obtained for the studied samples after six cycles in 0.1 M NaOH + 0.1 M KOH solution.

The EIS was also employed to assess the electrochemical response of the oxide films generated in alkaline conditions. As stated in the experimental section, the EIS measurements were performed in the passivity region every 50 mV by sweeping the potential anodically. Fig. 6 depicts some Bode diagrams that illustrate the evolution of the impedance as the potential is ennobling. The general tendency follows the same pattern in all samples, characterised by an impedance increase as the potential raise to around 0 $mV_{Hg/HgO}$; this can be associated with a gradual thickening of the oxide film. At higher potentials, the impedance decreases, probably owing to partial dissolution of the film. As expected, the AISI 304 L shows higher impedance values at all applied potentials, indicative of a better passive layer. On the other hand, the effect of the pre-stressing in the SMS samples can be appreciated since the impedance decrease of the s-SMS samples seems more pronounced than that of the u-SMS .

The obtained spectra are modelled using the electrical equivalent circuit shown in Fig. 6(d) which consists of a hierarchical arrangement of two RC elements. The good agreement between the experimental and fitted data can corroborate in Fig. 6(a) to (c). In this model, $R_e$ accounts for the electrolyte resistance, the high-frequency time constant, $R_{ct}C_{dl}$, is related to the charge transfer resistance and the double layer capacitance at the oxide layer-electrolyte interface. Considering the semi-conductive character of this film, the double layer capacitance should embrace the contributions of the space charge capacitance, $C_{SC}$; and the Helmotz layer capacitance, $C_H$, thus the overall $C_{dl}$ can be considered a combination of these two components in series [39]. The low-frequency time constant, $R_{redox}C_{redox}$, can be associated with the redox processes in the oxide layer. Both capacitances ($C_{dl}$ and $C_{redox}$) are affected by the corresponding Cole-Cole type dispersion coefficient, $\alpha_i$. A similar interpretation has been made by other researchers [27,30]. Fig. 7 displays the evolution of the described parameters with the applied potential.

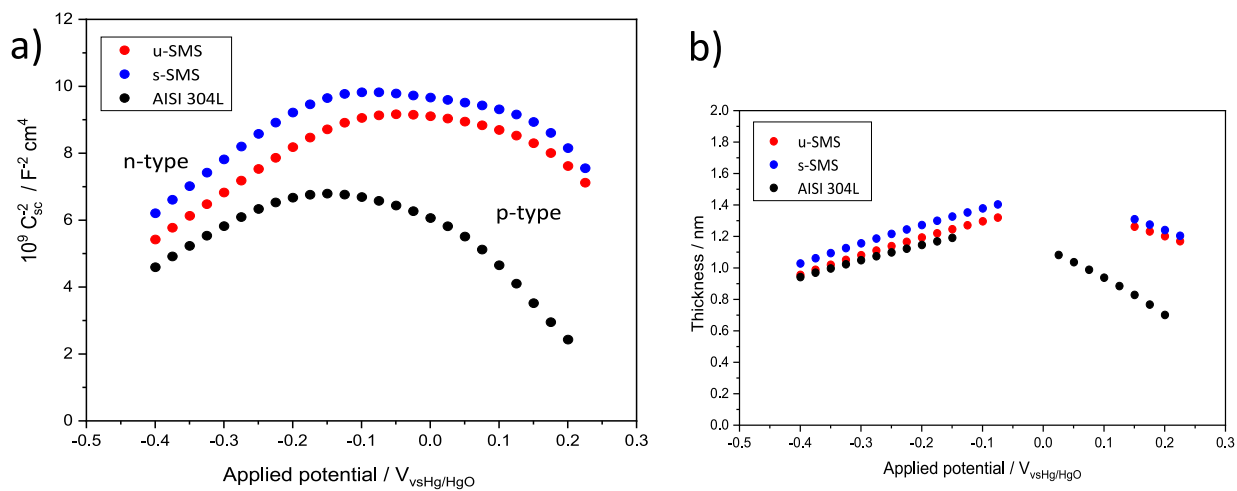

**Fig. 5.** M-S plots for the passive films generated on the studied samples after CVs (six cycles) in NaOH 0.1M + KOH 0.1 M solution (a), and the thickness of the space-charge layer, $L_{SC}$, as a function of the applied potential (b).

**Table 2**
Donor ($N_D$) and acceptor ($N_A$) densities extracted from the M-S plots of the generated oxide films.

| Alloy | $N_D$ (cm$^{-3}$) | $N_A$ (cm$^{-3}$) |
| --- | --- | --- |
| u-SMS | $6.7 \times 10^{20}$ | $5.3 \times 10^{20}$ |
| s-SMS | $6.1 \times 10^{20}$ | $4.9 \times 10^{20}$ |
| AISI 304L | $8.0 \times 10^{20}$ | $4.5 \times 10^{20}$ |

The $C_{dl}$ evolution shows an initial period characterized by nearly constant values, consistent with the low current density recorded in the CVs (see Fig. 3). At higher potentials, the capacitance rises markedly, mainly for the 304 L. As aforementioned, it should be related to the onset of $Cr^{6+}$ and $Mn^{3+}$ species formation, which could indicate a film thinning. However, Sudesh et al. [34] suggest instead that this rise in the capacitance probably reflects an increase in the film's conductivity owing to the rise in electron hole concentration, characteristic of p-type semiconductor behaviour. This interpretation agrees with the higher $C_{dl}$ increase observed for the 304 L. Parallel evolution is observed in the $R_{ct}$ values, which keep approximately constant, as expected for the passivity region. Only at the higher potentials a slight diminution is appreciated,

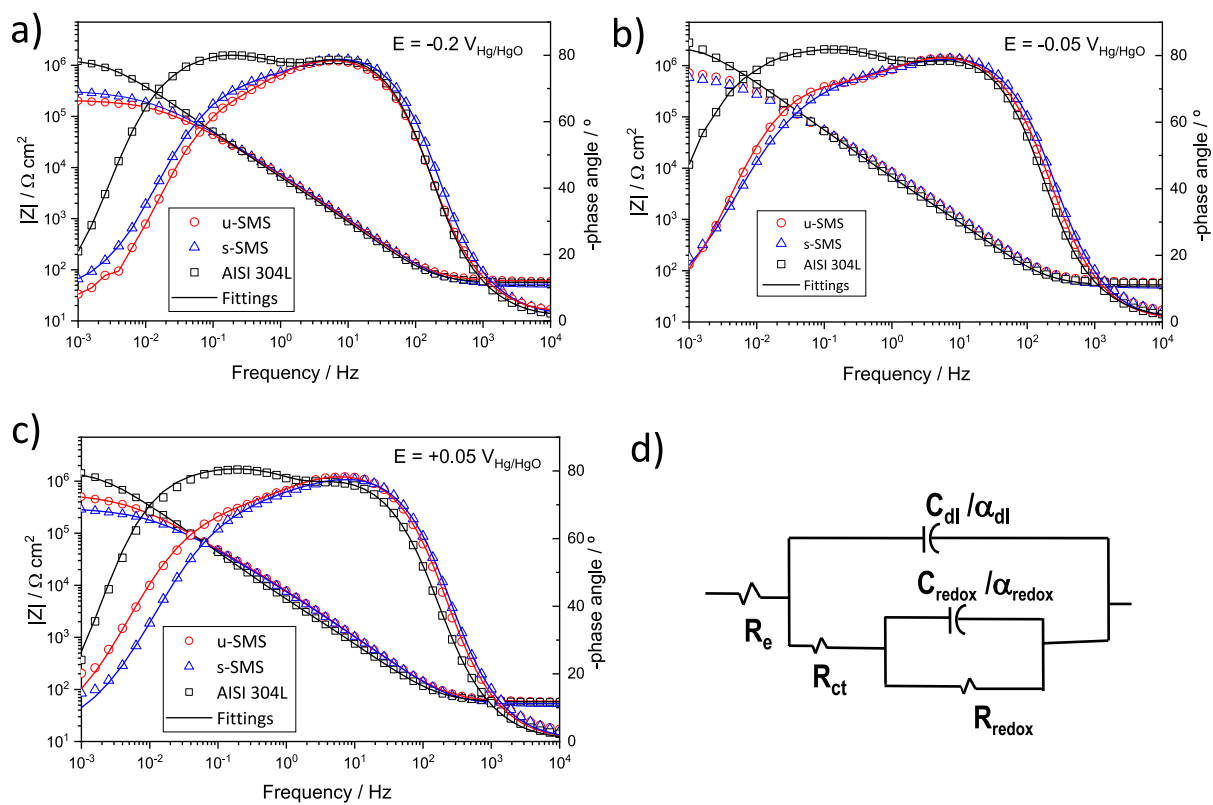

**Fig. 6.** Measured and fitted Bode plots for the studied samples in NaOH 0.1 $M$ + KOH 0.1 M solution at different potentials: (a) $-0.2V_{Hg/HgO}$, (b) $-0.05V_{Hg/HgO}$, (c) $+0.05V_{Hg/HgO}$, and (d) the electrical equivalent circuit used to fit the experimental impedance data.

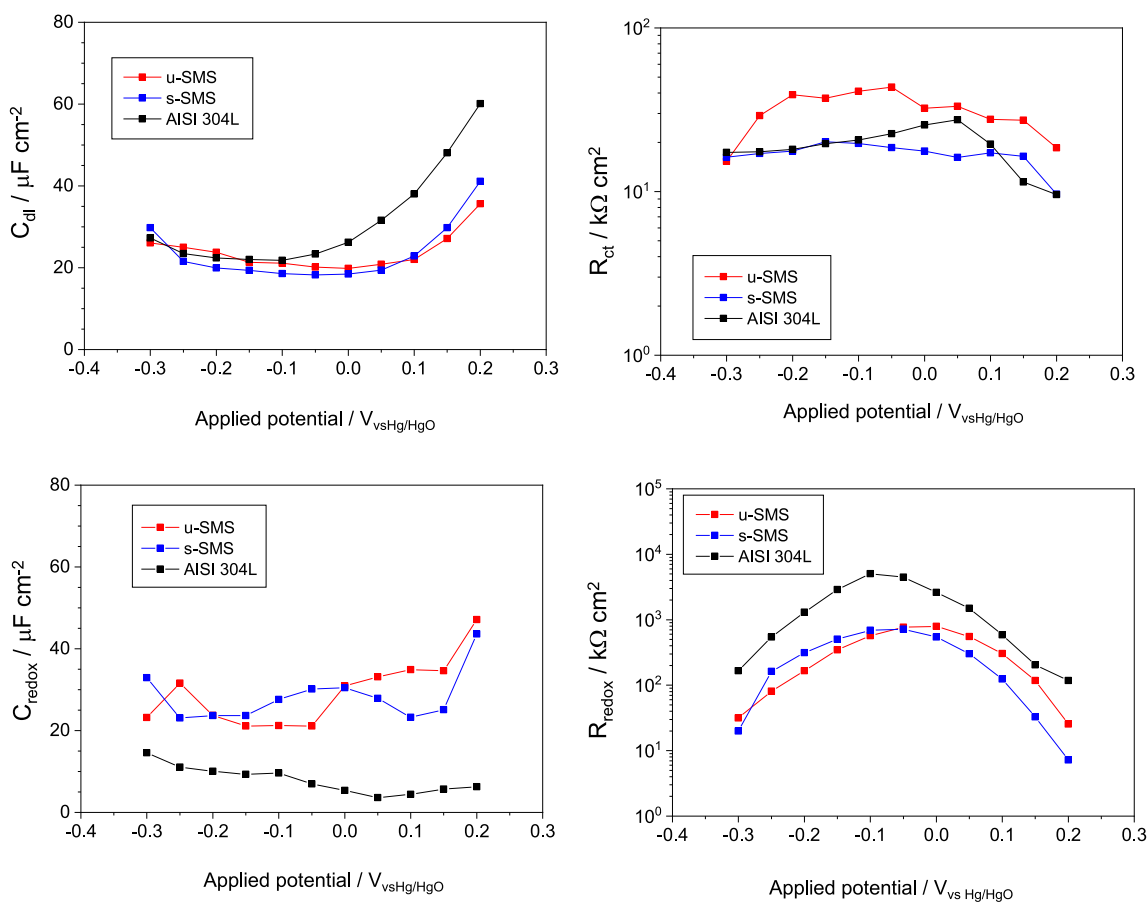

**Fig. 7.** Evolution with the applied potential of the high frequency ($R_{ct}C_{dl}$) and low frequency ($R_{redox}C_{redox}$) time constants corresponding to the electrical equivalent circuit depicted in Fig. 6d (the physical meaning is explained in the text).

which agrees with the capacitance increasing.

The low-frequency time constant provides information about the redox activity in the oxide films. The capacitance values are small, confirming this region's low oxidation rate. Despite that, the redox activity is higher for SMS samples and tends to increase at more anodic potentials, reflecting the Mn activity previously mentioned. The resistance evolution follows the reciprocal of the voltammograms profile. The initial increase indicates a slowdown in the redox activity, probably because as the thickness of the film increases, it adopts a more stable structure. At higher applied potentials, a continuous decrease is observed, which supports the activation of the Mn activity observed in the capacitance evolution. There are slight differences depending on the state. Thus, the resistance reduction is more substantial for pre-strained SMS samples. In the case of AISI 304 L, the resistance decreasing is not correlated to the capacitance evolution. Although it is not fully understood, it could result from increased film conductivity.

### 3.3. Electrochemical behaviour of the passive samples in aggressive media

The stability of the passive films generated in alkaline conditions can be compromised by aggressive ions, such as chlorides, or by the loss of alkalinity due to the carbonation phenomenon. Therefore, it is worth knowing the behaviour of the passivated SMSs in these aggressive environments. Fig. 8(a) depicts the polarization curves of the studied samples immersed in 0.1 M NaOH + 0.1 M KOH + 0.1 M NaCl solution. The passive films were previously generated in chloride-free alkaline conditions. As can be seen, the presence of chlorides has no significant effect on the stability of the oxide films. The slightly anodic values registered for u-SMS and s-SMS compared to AISI 304 L can be due to the thicker oxide films generated in these systems, as the XPS results corroborate. In the forward curve, a continuous increase in the current density is appreciated. However, the low recorded values confirm the passive state of the samples.

A more detailed analysis of the reverse curves reveals some changes if they are compared with those without chlorides. Fig. 8(b) shows a detail of these curves. In all samples, the potential at zero current is shifted anodically, which indicates that the films are enriched in the oxidised species ($Fe^{3+}$ and $Mn^{3+}/Mn^{4+}$). This effect is especially significant in s-SMS. In addition, the zero current potential observed at the more anodic values in the AISI 304 L ($E = 0.422\ V_{SCE}$) is not observed in the presence of chlorides, maybe owing to the formation of $NiCl_2$ that is a water-soluble species, the absence of Ni oxides in the presence of chlorides had been demonstrated in previous works [30]. These results indicate that, even though the passive layers withstand a 0.1 M chloride concentration, these ions induce surface modifications in the oxide films, which are particularly remarkable in the s-SMS.

The same tendency is observed at pH = 12.5 and pH = 12 (not included here). The passive films remain intact in all the studied cases. At lower pH values, distinct behaviour is recorded depending on the type of samples, as Fig. 9(a) to (c) illustrate.

The cyclic voltammograms recorded at pH = 11.5 (Fig. 9a) are characterised by an anodic shifting of the corrosion potential of the SMSs compared to the 304 L. A more anodic potential corresponds to a more oxidised layer which in turn corresponds to a more dehydrated layer (decreased local [$OH^-$]), facilitating the $Cl^-$ access to the passive film [30]. The higher value corresponds to the s-SMS; therefore, it will be prone to chloride attack, as corroborates the passivity breakdown observed at a potential of 0.478 $V_{SCE}$. In the reverse curve, there is no zero current potential at values nobler than the corrosion potential, which means the passive film is not re-passivated. On the other hand, the u-SMS and 304 L samples remain in a passive state. At pH = 11, the passivity breakdown for u-SMS samples is observed ($E = 0.328$ V$_{SCE}$). Moreover, a shortening in the passivity domain of s-SMS samples is appreciated ($E = 0.308$ V$_{SCE}$), corresponding to an increase of the [Cl$^-$]/[OH$^-$] ratio. As it was observed for s-SMS at pH = 11.5, no re-passivation takes place in neither s-SMS nor u-SMS.

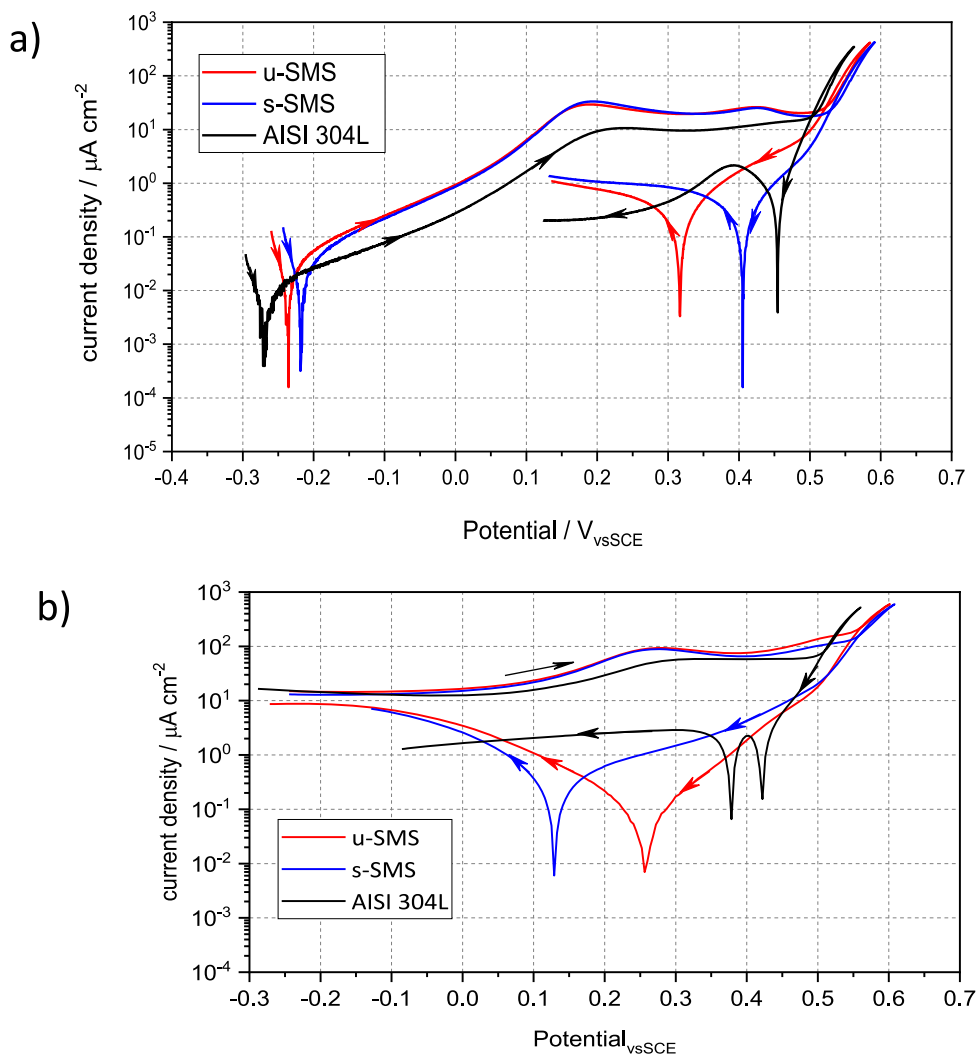

**Fig. 8.** Cyclic voltammograms obtained for the studied samples after passivated and immersed in 0.1 M NaOH + 0.1 M KOH + 0.1 M NaCl solution (pH = 13) at 1mV·s$^{-1}$ scan rate (a), a detail of the cyclic voltammogram obtained in 0.1 M NaOH + 0.1 M KOH solution (6th cycle) using 5mV·s$^{-1}$ scan rate.

Fig. 9(c) shows the behaviour of the shape memory steels in contrast to the AISI 304 L at pH = 10. This high [Cl$^-$]/[OH$^-$] ratio leads to an active film dissolution, with no passivation region regardless of the SMS state. On the contrary, AISI 304 L remains passive in these conditions. Pitting corrosion in this material is observed at pH = 9.5 at a potential of 0.418 V$_{SCE}$, although metastable pits are appreciated at lower potentials. The reverse curve exhibits the re-passivation potential at $E_{rp} = 0.310$ V$_{SCE}$, which suggests the regeneration of the passive film.

The higher resistance to pitting corrosion of AISI 304 L is expected, and was already observed by other researchers [30,40,41]. The higher susceptibility to the localized corrosion of the SMS is understandable considering the presence of elements like Mn, necessary to attain the shape memory effect, which is harmful to the corrosion resistance [42–44]. Despite that, they can generate a passive film in alkaline conditions that withstand the presence of chlorides at lower pHs than mild carbon steel [45]. This better passive behaviour should be due to the presence of Cr.

On the other hand, although the corrosion resistance of the SMS is similar in both states, the pre-strained SMS exhibits slightly lower resistance to chloride depassivation, in agreement with the results obtained by other researchers [14]. Although the reason is not well-understood, Söderberg et al. suggest that it can be due to the high interface area between the two phases (γ-austenite and ε-martensite) after deformation, which promotes more active zones to metal dissolution [8]. Additionally, the generation of the mechanical twins (see Fig. 1b) in the s-SMS could contribute to an increase in its surface energy.

Fig. 10 depicts the corroded surfaces after the passive film breakdown. In the SMS samples, the same morphology is observed in both states, characterised by a localised attack at the grain boundary. As stated in Section 3.1 (Fig. 1), this area corresponds to the δ-ferrite phase obtained during the solidification process. These results are in good agreement with previous research that shows the increase of the sensitivity to the intergranular corrosion of austenitic steels due to the presence of δ-ferrite [46–48]. Regarding the 304 L samples, those exhibit few pits with a size of about 150–200 μm and with the morphology usually observed in the stainless steels, i.e. they tend to grow under broken lacy covers [49,50].

## 4. Conclusions

The present work approached the characterization of a Fe-17Mn-6Si-

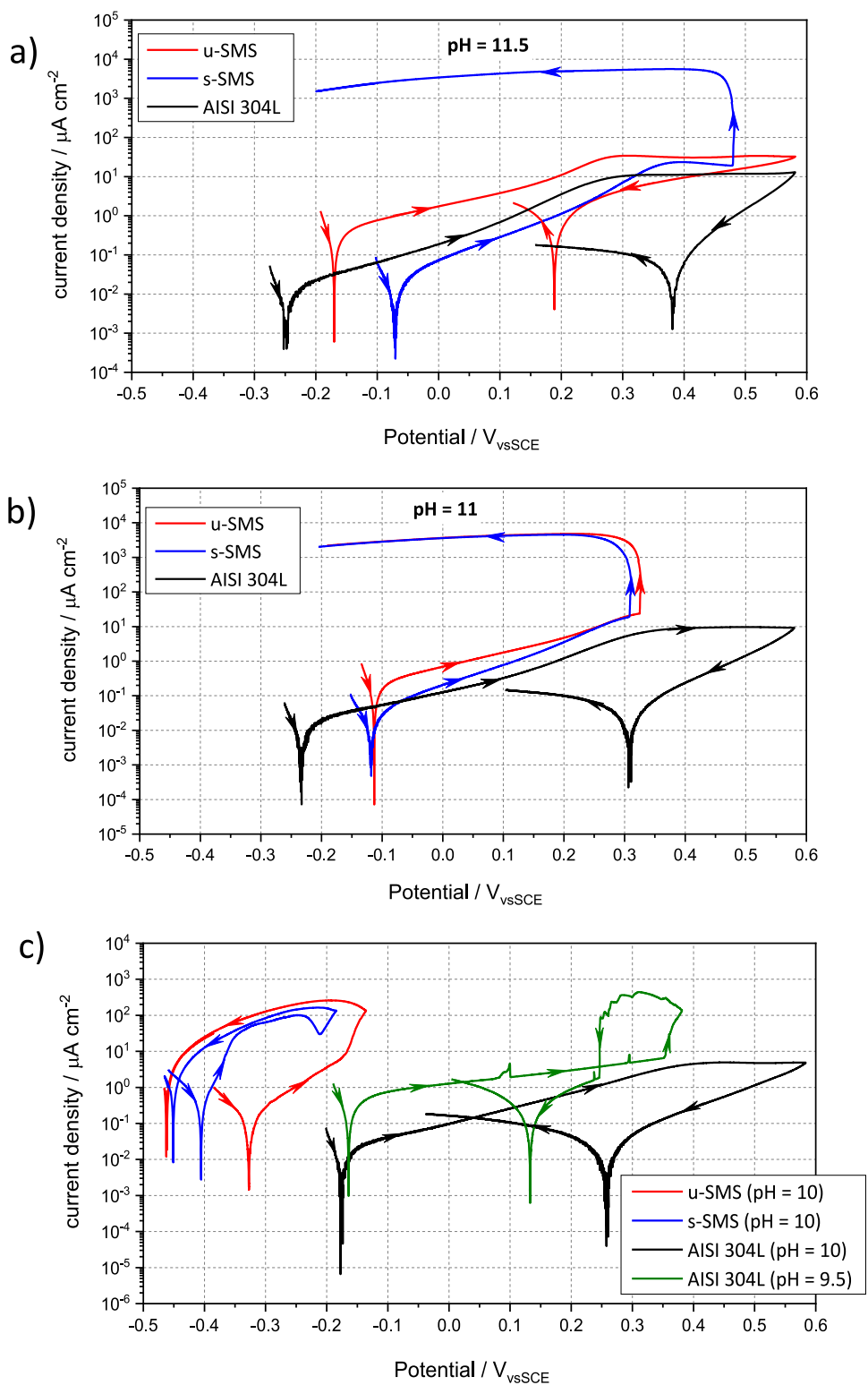

**Fig. 9.** Polarization curves recorded for u-SMS, s-SMS and AISI 304 L samples at different pHs maintaining a constant chloride concentration of 0.1 M.

10Cr-4Ni-1(V,C) shape memory alloy in unstrained and 5% pre-strained conditions. The study was performed in alkaline conditions close to the concrete pores' solution. The main results can be summarized in the following conclusions:

- The morphological characterisation revealed the presence of γ-austenite and ε-stress-induced martensite, in both states. In addition, a new δ-ferrite phase was also observed, located at grain boundaries.

- By cyclic voltammetry, a passive film was generated in alkaline conditions, not only on SMS samples but also on AISI 304 L, for comparative purposes. The CVs of SMS presented some differences compared with the AISI 304 L that should be associated with the presence of Mn. The most significant difference was related to the lower current density recorded at the magnetite formation peak in

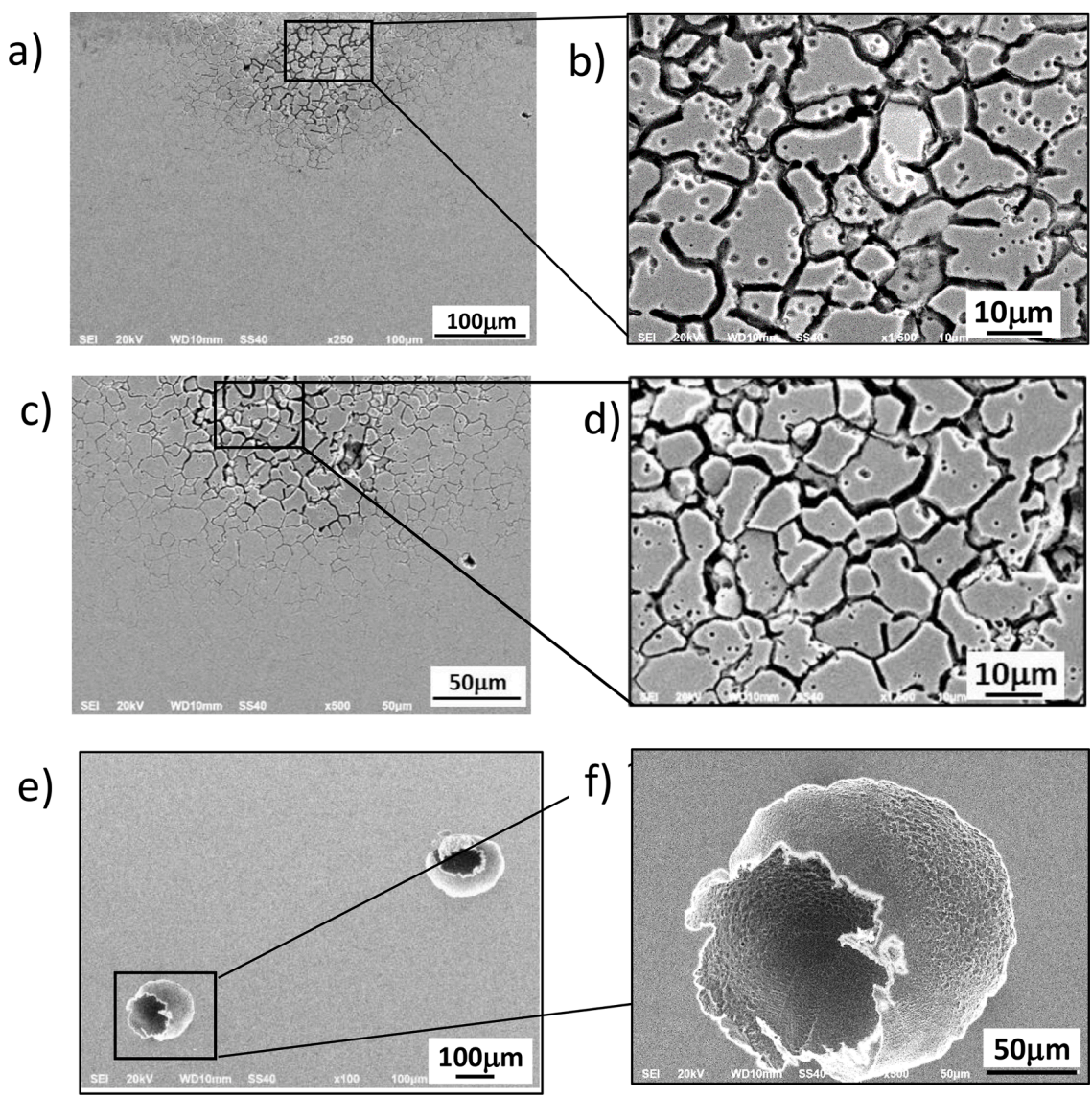

**Fig. 10.** SEM images of the metallic surfaces after the passive film breakdown at different magnifications: a general overview of u-SMS (a), s-SMS (c) and AISI 304 L (e), and a detail of the localised attack morphology of u-SMS (b), s-SMS (d) and AISI 304 L (f).

- the case of the SMS, which may be associated with the presence of manganese oxides that partially blocked the iron activity.
- The XPS results indicated that the thickness of the films was around 20 nm and corroborated the presence of oxides/hydroxides of Fe, Cr, Mn and Ni in the oxide film.
- The electronic characteristics of these films revealed a dual response, n-type semiconducting behaviour at lower potentials and p-type at more anodic ones.
- The impedance measurements performed towards the anodic direction initially reflected a gradual thickening of the oxide films. However, at higher potentials, there was some dissolution. In addition, differences were observed in the impedance values depending on the state of the SMS; the lower values registered for the pre-strained SMS samples indicated the generation of less stable oxide film, probably due to the stressing procedure.
- These previous conclusions were corroborated when the samples were immersed in aggressive conditions, involving the presence of chlorides and the loss of alkalinity. The SMS was more sensitive to the decreasing pH and chloride ions presence than the AISI 304 L, mainly in the pre-strained stage. This result may limit the use of this SMS in pre-stressing reinforcement components.
- The corroded surface morphology observed in SMS samples revealed the intergranular corrosion due to the δ-ferrite phase. Therefore, in this line, the optimization of the production procedure of this alloy to avoid the δ-ferrite formation should increase its corrosion resistance.

**CRediT authorship contribution statement**

**A. Collazo:** Visualization, Validation. **R. Figueroa:** Investigation, Methodology. **C. Mariño-Martínez:** Resources. **X.R. Nóvoa:** Writing – review & editing, Supervision. **C. Pérez:** Conceptualization, Writing – original draft.

**Declaration of Competing Interest**

The corresponding authors on behalf of all authors declare any financial or any potential competing or non-financial interests of the manuscript.

**Data availability**

Data will be made available on request.

**Acknowledgements**

The authors gratefully acknowledge the financial support for this work by the National Program of Spanish Ministry of Science and Innovation under the contract TED2021–130605B-I00.

**References**

[1] F.C. Nascimento-Borges, Iron based shape memory alloys: mechanical and structural properties, shape memory alloys - processing, characterization and applications, Shape Mem. Alloy. Process. Charact. Appl. (2013) 116–128, https://doi.org/10.5772/51877.

[2] A. Sato, E. Chishima, K. Soma, T. Mori, Shape memory effect in $\gamma \rightleftarrows \varepsilon$ transformation in Fe-30Mn-1Si alloy single crystals, Acta Metall. 30 (1982) 1177–1183, https://doi.org/10.1016/0001-6160(82)90011-6.

[3] W.J. Lee, B. Weber, G. Feltrin, C. Czaderski, M. Motavalli, C. Leinenbach, Phase transformation behavior under uniaxial deformation of an Fe-Mn-Si-Cr-Ni-VC shape memory alloy, Mater. Sci. Eng. A 581 (2013) 1–7, https://doi.org/10.1016/j.msea.2013.06.002.

[4] A. Cladera, B. Weber, C. Leinenbach, C. Czaderski, M. Shahverdi, M. Motavalli, Iron-based shape memory alloys for civil engineering structures: an overview, Constr. Build. Mater. 63 (2014) 281–293, https://doi.org/10.1016/j.conbuildmat.2014.04.032.

[5] W.J. Lee, B. Weber, G. Feltrin, C. Czaderski, M. Motavalli, C. Leinenbach, Stress recovery behaviour of an Fe–Mn–Si–Cr–Ni–VC shape memory alloy used for prestressing, Smart Mater. Struct. 22 (2013), 125037, https://doi.org/10.1088/0964-1726/22/12/125037.

[6] M. Shahverdi, C. Czaderski, M. Motavalli, Iron-based shape memory alloys for prestressed near-surface mounted strengthening of reinforced concrete beams, Constr. Build. Mater. 112 (2016) 28–38, https://doi.org/10.1016/j.conbuildmat.2016.02.174.

[7] H. Otsuka, H. Yamada, T. Maruyama, H. Tanahashi, S. Matsuda, M. Murakami, Effects of alloying additions on Fe-Mn-Si shape memory alloys, ISIJ Int. 30 (1990) 674–679, https://doi.org/10.2355/isijinternational.30.674.

[8] O. Söderberg, X.W. Liu, P.G. Yakovenko, K. Ullakko, V.K. Lindroos, Corrosion behaviour of Fe-Mn-Si based shape memory steels trained by cold rolling, Mater. Sci. Eng. A 273–275 (1999) 543–548, https://doi.org/10.1016/S0921-5093(99)00396-2.

[9] S. Kajiwara, D. Liu, T. Kikuchi, N. Shinya, Remarkable improvement of shape memory effect in Fe-Mn-Si based shape memory alloys by producing NbC precipitates, Scr. Mater. 44 (2001) 2809–2814, https://doi.org/10.1016/S1359-6462(01)00978-2.

[10] H. Kubo, K. Nakamura, S. Farjami, T. Maruyama, Characterization of Fe-Mn-Si-Cr shape memory alloys containing VN precipitates, Mater. Sci. Eng. A 378 (2004) 343–348, https://doi.org/10.1016/j.msea.2003.10.359.

[11] M.J. Lai, Y.J. Li, L. Lillpopp, D. Ponge, S. Will, D. Raabe, On the origin of the improvement of shape memory effect by precipitating VC in Fe–Mn–Si-based shape memory alloys, Acta Mater. 155 (2018) 222–235, https://doi.org/10.1016/j.actamat.2018.06.008.

[12] H.C. Lin, K.M. Lin, C.S. Lin, T.M. Ouyang, The corrosion behavior of Fe-based shape memory alloys, Corros. Sci. 44 (2002) 2013–2026, https://doi.org/10.1016/S0010-938X(02)00027-6.

[13] B.C. Maji, C.M. Das, M. Krishnan, R.K. Ray, The corrosion behaviour of Fe-15Mn-7Si-9Cr-5Ni shape memory alloy, Corros. Sci. 48 (2006) 937–949, https://doi.org/10.1016/j.corsci.2005.02.024.

[14] W.J. Lee, R. Partovi-Nia, T. Suter, C. Leinenbach, Electrochemical characterization and corrosion behavior of an Fe-Mn-Si shape memory alloy in simulated concrete pore solutions, Mater. Corros. 67 (2016) 839–846, https://doi.org/10.1002/maco.201508701.

[15] R. Babić, M. Metikoš-Huković, Semiconducting properties of passive films on AISI 304 and 316 stainless steels, J. Electroanal. Chem. 358 (1993) 143–160, https://doi.org/10.1016/0022-0728(93)80435-K.

[16] P. Lu, S. Sharifi-Asl, B. Kursten, D.D. Macdonald, The irreversibility of the passive state of carbon steel in the alkaline concrete pore solution under simulated anoxic conditions, J. Electrochem. Soc. 162 (2015) C572–C581, https://doi.org/10.1149/2.0731510jes.

[17] K. Azumi, T. Ohtsuka, N. Sato, Mott-schottky plot of the passive film formed on iron in neutral borate and phosphate solutions, J. Electrochem. Soc. 134 (1987) 1352–1357, https://doi.org/10.1149/1.2100672.

[18] J.C. Lippold, D.J. Kotecki, Welding Metallurgy and Weldability of Stainless Steel, John Wiley & Sons, Inc., 2005.

[19] K.L. Chao, H.Y. Liao, J.J. Shyue, S.S. Lian, Corrosion behaviour of high nitrogen nickel-free Fe-16Cr-Mn-Mo-N stainless steel, Metall. Mater. Trans. B 45B (2014) 381.

[20] B.C. Maji, M. Krishnan, V.V. Rama Rao, The microstructure of an Fe-Mn-Si-Cr-Ni stainless steel shape memory alloy, Metall. Mater. Trans. A Phys. Metall. Mater. Sci. 34 A (2003) 1029–1042, https://doi.org/10.1007/s11661-003-0124-y.

[21] Y. Wen, H. Peng, C. Wang, Q. Yu, N. Li, A novel training-free cast Fe-Mn-Si-Cr-Ni shape memory alloy based on formation of martensite in a domain-specific manner, Adv. Eng. Mater. 13 (2011) 48–56, https://doi.org/10.1002/adem.201000200.

[22] N. Stanford, D.P. Dunne, Thermo-mechanical processing and the shape memory effect in an Fe-Mn-Si-based shape memory alloy, Mater. Sci. Eng. A 422 (2006) 352–359, https://doi.org/10.1016/j.msea.2006.02.009.

[23] C.M. Abreu, M.J. Cristóbal, R. Losada, X.R. Nóvoa, G. Pena, M.C. Pérez, High frequency impedance spectroscopy study of passive films formed on AISI 316 stainless steel in alkaline medium, J. Electroanal. Chem. 572 (2004) 335–345, https://doi.org/10.1016/j.jelechem.2004.01.015.

[24] L. Freire, M.A. Catarino, M.I. Godinho, M.J. Ferreira, M.G.S. Ferreira, A.M. P. Simões, M.F. Montemor, Electrochemical and analytical investigation of passive films formed on stainless steels in alkaline media, Cem. Concr. Compos. 34 (2012) 1075–1081, https://doi.org/10.1016/j.cemconcomp.2012.06.002.

[25] S. Haupt, H.H. Strehblow, Corrosion, layer formation, and oxide reduction of passive iron in alkaline solution: a combined electrochemical and surface analytical study, Langmuir 3 (1987) 873–885, https://doi.org/10.1021/la00078a003.

[26] C.M. Abreu, M.J. Cristóbal, R. Losada, X.R. Nóvoa, G. Pena, M.C. Pérez, The effect of Ni in the electrochemical properties of oxide layers grown on stainless steels, Electrochim. Acta 51 (2006) 2991–3000, https://doi.org/10.1016/j.electacta.2005.08.033.

[27] B. Messaoudi, S. Joiret, M. Keddam, H. Takenouti, Anodic behaviour of manganese in alkaline medium, Electrochim. Acta 46 (2001) 2487–2498, https://doi.org/10.1016/S0013-4686(01)00449-2.

[28] Y.S. Zhang, X.M. Zhu, S.H. Zhong, Effect of alloying elements on the electrochemical polarization behavior and passive film of Fe-Mn base alloys in various aqueous solutions, Corros. Sci. 46 (2004) 853–876, https://doi.org/10.1016/j.corsci.2003.09.002.

[29] T. Yamashita, P. Hayes, Analysis of XPS spectra of $Fe^{2+}$ and $Fe^{3+}$ ions in oxide materials, Appl. Surf. Sci. 254 (2008) 2441–2449, https://doi.org/10.1016/j.apsusc.2007.09.063.

[30] C.M. Abreu, M.J. Cristóbal, R. Losada, X.R. Nóvoa, G. Pena, M.C. Pérez, Long-term behaviour of AISI 304L passive layer in chloride containing medium, Electrochim. Acta 51 (2006) 1881–1890, https://doi.org/10.1016/j.electacta.2005.06.040.

[31] A. Di Paola, Semiconducting properties of passive films on stainless steels, Electrochim. Acta 34 (1989) 203–210, https://doi.org/10.1016/0013-4686(89)87086-0.

[32] N.E. Hakiki, M. Da Cunha Belo, A.M.P. Simões, M.G.S. Ferreira, Semiconducting properties of passive films formed on stainless steels: influence of the alloying elements, J. Electrochem. Soc. 145 (1998) 3821–3829, https://doi.org/10.1149/1.1838880.

[33] Y. Zhang, H. Luo, Q. Zhong, H. Yu, J. Lv, Characterization of passive films formed on as-received and sensitized AISI 304 stainless steel, Chin. J. Mech. Eng. Engl. Ed. 32 (2019), https://doi.org/10.1186/s10033-019-0336-8.

[34] T.L. Sudesh L. Wijesinghe, D.J. Blackwood, Photocurrent and capacitance investigations into the nature of the passive films on austenitic stainless steels, Corros. Sci. 50 (2008) 23–34, https://doi.org/10.1016/j.corsci.2007.06.009.

[35] X. Feng, X. Lu, Y. Zuo, D. Chen, The passive behaviour of 304 stainless steels in saturated calcium hydroxide solution under different deformation, Corros. Sci. 82 (2014) 347–355, https://doi.org/10.1016/j.corsci.2014.01.039.

[36] C.A. Della Rovere, J.H. Alano, R. Silva, P.A.P. Nascente, J. Otubo, S.E. Kuri, Characterization of passive films on shape memory stainless steels, Corros. Sci. 57 (2012) 154–161, https://doi.org/10.1016/j.corsci.2011.12.022.

[37] M. Mandel, V. Kietov, R. Hornig, M. Vollmer, J.M. Frenck, C. Wüstefeld, D. Rafaja, T. Niendorf, L. Krüger, On the polarisation and Mott-Schottky characteristics of an Fe-Mn-Al-Ni shape-memory alloy and pure Fe in NaCl-free and NaCl-contaminated $Ca(OH)_2$,sat solution–a comparative study, Corros. Sci. (2021) 179, https://doi.org/10.1016/j.corsci.2020.109172.

[38] N.B. Hakiki, S. Boudin, B. Rondot, M. Da Cunha Belo, The electronic structure of passive films formed on stainless steels, Corros. Sci. 37 (1995) 1809–1822, https://doi.org/10.1016/0010-938X(95)00084-W.

[39] J. Huang, R. Qiao, G. Feng, B.G. Sumpter, V. Meunier, Modern theories of carbon-based electrochemical capacitors, in: F. Béguin, E. Frqckowiak (Eds.), Supercapacitors: Materials, Systems, and Applications, Wiley-VCH GmbH¬Co. KGaA, 2013, pp. 167–206, https://doi.org/10.1002/9783527646661.ch5.

[40] L. Freire, M.J. Carmezim, M.G.S. Ferreira, M.F. Montemor, The electrochemical behaviour of stainless steel AISI 304 in alkaline solutions with different pH in the presence of chlorides, Electrochim. Acta 56 (2011) 5280–5289, https://doi.org/10.1016/j.electacta.2011.02.094.

[41] A.A. Dastgerdi, A. Brenna, M. Ormellese, M.P. Pedeferri, F. Bolzoni, Experimental design to study the influence of temperature, pH, and chloride concentration on the pitting and crevice corrosion of UNS S30403 stainless steel, Corros. Sci. 159 (2019), 108160, https://doi.org/10.1016/j.corsci.2019.108160.

[42] K.J. Park, H.S. Kwon, Effects of Mn on the localized corrosion behavior of Fe-18Cr alloys, Electrochim. Acta 55 (2010) 3421–3427, https://doi.org/10.1016/j.electacta.2010.01.006.

[43] C.A. Della Rovere, J.H. Alano, R. Silva, P.A.P. Nascente, J. Otubo, S.E. Kuri, Influence of alloying elements on the corrosion properties of shape memory stainless steels, Mater. Chem. Phys. 133 (2012) 668–673, https://doi.org/10.1016/j.matchemphys.2012.01.049.

[44] H.Y. Ha, M.H. Jang, T.H. Lee, Influences of Mn in solid solution on the pitting corrosion behaviour of Fe-23 wt%Cr-based alloys, Electrochim. Acta 191 (2016) 864–875, https://doi.org/10.1016/j.electacta.2016.01.118.

[45] M. Liu, X. Cheng, X. Li, T.J. Lu, Corrosion behavior of low-Cr steel rebars in alkaline solutions with different pH in the presence of chlorides, J. Electroanal. Chem. 803 (2017) 40–50, https://doi.org/10.1016/j.jelechem.2017.09.016.

[46] T.M. Devine, Mechanism of intergranular corrosion and pitting corrosion of austenitic and duplex 308 stainless steel, J. Electrochem. Soc. 126 (1979) 374–385, https://doi.org/10.1149/1.2129046.

[47] G. Bai, S. Lu, D. Li, Y. Li, Intergranular corrosion behavior associated with delta-ferrite transformation of Ti-modified Super304H austenitic stainless steel, Corros. Sci. 90 (2015) 347–358, https://doi.org/10.1016/j.corsci.2014.10.031.

[48] R. Mohammed, G. Madhusudhan Reddy, K. Srinivasa Rao, Microstructure and pitting corrosion of shielded metal arc welded high nitrogen stainless steel, Def. Technol. 11 (2015) 237–243, https://doi.org/10.1016/j.dt.2015.04.002.

[49] G.T. Burstein, P.C. Pistorius, S.P. Mattin, The nucleation and growth of corrosion pits on stainless steel, Corros. Sci. 35 (1993) 57–62.

[50] P. Ernst, R.C. Newman, Pit growth studies in stainless steel foils. I. Introduction and pit growth kinetics, Corros. Sci. 44 (2002) 927–941, https://doi.org/10.1016/S0010-938X(01)00133-0.